\documentclass{article}

\input{tcilatex}

\begin{document}

\subsection{\textbf{Experimental Evidence Of The Strong Phonon Scattering In
Isotopical Disordered Systems: The Case LiH}$_{\mathbf{x}}$\textbf{D}$_{%
\mathbf{1-x}}$\textbf{.}}

\bigskip

\ \ \ \ \ \ \ \ \ \ \ \ \ \ \ \ \ \ \ \ \ \ \ \ \ \ \ \ \ \ \ \ \ \ \ \ \ \
\ \ \ \ \ \ \ \ \ \textbf{V.G. Plekhanov}

\textbf{Fonoriton Science Laboratory, Garon Ltd., PO Box 2632, Tallinn-38,
EEOO38, ESTONIA}

\bigskip

\textbf{\ }

\textbf{Abstract.} We present Raman spectra of LiH$_{x}$D$_{1-x}$ mixed
crystals at room temperature and in a wide range of isotope compositions (0 
\TEXTsymbol{<} x \TEXTsymbol{<}1). The line broadening and energy shift of
the Raman lines for the optical phonons were investigated for the first
time. The observation of the local mode vibration, the two-mode behaviour of
the LO ($\Gamma $) phonons at large isotope concentration as well as large
lines broadening directly evidence the strong additional phonon scattering
due to the isotope-induced disorder. The obtained results could not describe
in the weak phonon scattering approximation of the CPA model.

\bigskip

Some of the most interesting and important effects in the atomic vibrations
of solids arise from the presence of small concentration of impurity
(defects) in the perfect lattice. The important effects of the point defects
are the following. Firstly, the addition of a small impurity content may
destroy the translational symmetry of the unperturbed perfect crystal in
addition to slight change of its lattice modes. Such changes are often
sufficient to allow interaction with light radiation in cases where it would
not occur otherwise (absorption process etc.). Secondly, the impurity atoms
often introduce special modes into the phonon spectrum which are not
characteristic for the host lattice as a whole but rather that for the
atomic environment surrounding the impurity. These special vibrational modes
are called local modes, gap modes and resonant band modes (see, e.g. [1,2]).
As it is well-known , a localised mode is characterized by its high
frequency which lies above the maximum vibrational frequency of the perfect
host crystal. The amplitude of this vibration is very large at the impurity
site and decreases ASH very rapidly with increasing distance from the
impurity. A gap mode has a frequency which lies in the gap between the
optical and acoustic bands. If the impurity content of a solid is increased
to such an extent that interaction between impurity atoms begins to play an
effective role, the systems should then be named a disordered solid rather
than an impure crystal.

The case is of particular interest when the impurity atoms are of the same
chemical nature, but have a different mass, i.e., the case of isotopic
impurities. In this case no deformation of the unit cells at the
substitutional sites occurs, and, since the crystal potential remains the
same as in the isotopically pure material, only the vibrational kinetic
energy is affected by the mass impurities. As it is well-known [1,2] the
lattice phonons in an atomic (ionic) crystal are true collective motions of
the atoms. The impurity atoms are involved in the collective motion, but,
owing to their different mass, they will follow the motion with a different
amplitude, thus giving rise to a change in the phase of the motion. The
impurities therefore play the role of scattering centers for the phonons and
their effect then can be described in terms of scattering processes. In the
case of optical phonons this will lead to a shift and to a broadening of the
phonon band in the Raman scattering spectra.

The effect of isotope substitution and its influence on the LiH lattice
dynamics has a long history. IR-absorption and \ - reflection [3-5], neutron
scattering [6] as well as Raman scattering [7-13] studies allow us to
discover the local [7] and gap [3] modes in the vibration spectra of LiH-D
mixed system. According to these studies the energy of the LO phonons at the 
$\Gamma $ point of the Brillouin zone is equal to 840 and 1100 cm$^{-1}$ for
LiD and LiH and for the local (LiD-H) and gap (LiH-D) modes - respectively
917 and 580 cm$^{-1}$. Moreover, the first paper [14] dealing with the
fundamental absorption, spectra of thin films of LiH and LiD showed that the
longwavelength (actually, exciton) peak of LiH is shifted towards
shortwavelength relative to that of LiH. This shift is 0.65 meV [14] at room
and 103 meV [15] at liquid helium temperature.

In recent years the effects of the isotopic substitution and disorder on the
phonon states of germanium [16,17] and diamond [18,19] have attracted
considerable interest, triggered by the availability of nearly isotopically
pure materials. There, it was shown that the results obtained on the shift
and broadening of the phonon lines could be accurately described at the
limit of the weak phonon scattering of the phonon coherent potential
approximation (CPA) model [20,21]. In the limit of weak phonon scattering
the small parameter of the theory is $\Delta $M/M \TEXTsymbol{<}\TEXTsymbol{<%
} 1 (where $\Delta $M is mass difference between the various isotopes and M
is average isotope mass), and the scattering potential due to the mass
defects is not strong enough to localize the $\Gamma $ point LO phonons.

In case of LiH-D mixed system the parameter $\Delta \mu $/$\mu \ast /$ $%
\rangle $ 1/2 (where $\mu $ is the reduced effective mass of the unit cell)
is not small. Therefore for this system we may expect to be at the other
extreme case, in which the scattering potential is quite strong. In this
letter we report the first results on the effects of the anharmonicity and
isotopic disorder on the vibronic properties of LiH$_{x}$D$_{1-x}$ (0 
\TEXTsymbol{<} x \TEXTsymbol{<} 1) mixed crystals, first of all concerning
the optical phonons. The optical method selected for investigation is the
classical Raman - scattering technique. In particular, not only will the
two-mode behaviour of the LO($\Gamma $) phonons at the large concentration
(x $\equiv $ 0.5) be shown, but also the direct connection of the shift and
broadening of their lines with a large anharmonicity as well as a strong
phonon scattering due to the isotopically induced disorder. Additionally,
our experimental results are direct evidence of the transition from local
mode vibration to the LO ($\Gamma $) phonons. The experimentally observed
asymmetry of the optical phonon lines is due to isotope-induced disorder as
well.

The apparatus used in the present study has been described elsewhere (Ref.
13 and 22 and the references cited therein). For convenience, however, we
shall briefly mention the principal points. Crystals were excited by various
argon laser lines (particularly 514.5, 488.0 and 467.5 nm) and also by the
second harmonic line of a YAG-Nd$^{3+}$ laser ($\lambda _{L}$ = 532 nm). In
all these cases the laser power was between 50 and 200 mW, depending on the
intensity of the scattered line which varied by more than one order of
magnitude. The Raman spectra were obtained using an automatic apparatus
based on a double-grating monochromator. The signal from this monochromator
was recorded with a cooled photomultiplier operating as a photon counter.
The results are stored in the memory of a personal computer, used also to
control the apparatus. The counting time usually did not exceed 10 s. A
typical slit width was less than 150 $\mu $m. To supress plasma fluorescence
an appropriate interference spike filter was used in the incident beam. The
90o scattering geometry was used most frequently, but similar results were
obtained also in backscattering configuration. Both LiH and mixed crystals
based on it were highly hygroscopic. Therefore, special measurements were
performed to monitor the state of the surface of our samples. Most
convenient method was to follow the peak in the Raman spectra at 1450 cm$%
^{-1}$, which represented the molecular vibrations of CO$_{2}$ [23] or a
peak at w = 500 cm$^{-1}$ due to OH vibrations [4,5]. Our samples were thick
plates freshly cleaved from high-quality bulk single crystals grown by a
modification of the Bridgman-Stockbarger methods (see [24,12,15] and the
literature cited therein). Apart from chemical and mass-spectroscopical
isotope concentration [12,24] their composition was verified also from the
energetic nomogram of the exciton ground state extracted from low
temperature reflectance and luminescence spectra. The last one is possible
as far as the exciton ground state reconstruction in the whole concentration
interval (0 \TEXTsymbol{<} x \TEXTsymbol{<} 1) possesses the one-mode
character (for details see Ref. 25).

\bigskip

Fig.1. Second-order Raman spectra of LiH$_{x}$D$_{1-x}$ mixed crystals: x =
0 (1); 0.05 (2); 0.12 (3) and 0.35 (4).

Fig. 1 shows a family of the Raman spectra of LiH$_{x}$D$_{1-x}$ mixed
crystals with relatively low hydrogen concentration (x \TEXTsymbol{<} 35\%).
Analogous spectra of the mixed crystals with a large isotope concentration
is depicted on the Fig. 2.

\bigskip Fig.2. Second-order Raman spectra of LiH$_{x}$D$_{1-x}$ mixed
crystals: x = 0 (1); 0.42 (2); 0.76 (3) and1.0 (4). The arrows indicated the
transition from LO($\Gamma $) phonon and local mode peaks of LiD crystal to
LO($\Gamma $) phonon peak of LiH crystal via an intermediate LO peaks of LiH$%
_{x}$D$_{1-x}$ mixed crystals.

\bigskip

First of all it is important to remark here on the close agreement between
the pictured spectra of virgin LiH and LiD and previously published data
[7-13]. Moreover, the practically constant low frequency structure of all
the spectra is remarkable, primarily associated with lithium ion vibrations
[9 -13,22]. The exception was the lowest frequency peak whose intensity
decreased on the increase in x and its origin is connected with the
substraction of the tranverse optical and acoustical phonons. As it was
shown previously [22], the practically linear nature of this dependence
could be used, in addition to other methods [25], as, an independent means
for determining the composition of LiH$_{x}$D$_{1-x}$ mixed crystals. On the
other hand, the high frequency part of the Raman spectra connected with
excitation of the optical phonons, changes considerably with the increase in
x.

Although from the exciton states nomogram [25] it can be concluded that the
investigated crystal (Fig. 1, curve 1) is pure LiD, a high frequency peak
can be clearly observed in the spectral region $\simeq $ 1850 cm$^{-1}$ (it
is more significant at the excitation with $\lambda _{L}$ = 532 nm (see,
also [13]). It must be mentioned that such a peak has not been observed in
pure LiH crystals. This peak has been observed before in the second-order
Raman spectra of LiD crystals and attributed, as mentioned above, to the
local vibration of hydrogen [7]. The peak observed in our spectra at $\simeq 
$1850 cm$^{-1}$ corresponds to 2$\omega _{loc}$. Although the actual
concentration of H in the used LiD crystal is very low and unknown, it can
certainly be accounted for the observed local mode in the second-order Raman
spectra. As we see, the increasing concentration causes the growth in the
local mode band intensity and a small shift of the frequency. This behaviour
remains up to x \TEXTsymbol{<} 10\% (for details see Ref. 10). These results
are in close agreement with the calculated concetration dependence of the
intensity and $\omega _{loc}$ of the local vibration (see, also Ref. 20,21).
With a further increase of x according to the results on the low temperature
exciton luminescence and resonant Raman spectra [15] the system consists of
mixed crystals. We should emphasize here that the successive investigation
of the isotope concentration dependence of the local vibration line's shape
allows us to measure the percolation threshold of these mixed crystals (for
details see, also Ref. 20,26 and references therein).

\qquad Returning to Fig. 2 in detail we examine the analysis of the Raman
spectra structure of LiH$_{x}$D$_{1-x}$ mixed crystals at the large isotope
concentration. With the further increase x \TEXTsymbol{>} 0.15 we observed
the decrease of the 2LO ($\Gamma $) phonon maximum intensity of the virgin
LiD crystal with a simultaneous growth of the higher frequency peak
intensity of the LiH$_{x}$D$_{1-x}$ mixed crystals. The last one was
attributed also to the renormalized LO ($\Gamma $) vibrations in the mixed
crystals (see, also Ref. 22). The final limit of this peak is the position
of LO ($\Gamma $) phonon peak of the virgin LiH crystal (Fig. 3).

\bigskip

\bigskip

Fig.3. Isotope composition dependence of the (a) linewidth (FWHM) and (b)
frequency of the optical phonons in the Raman spectra of LiH$_{x}$D$_{1-x}$
mixed crystals: a) 1, $\lambda _{excit}$ = 253.7 nm; 2 - $\lambda _{L}$ =
488.0 nm. \qquad \qquad b) 1 - results the present paper; 2, results on the
IR-absorption TO($\Gamma $) phonons in thin film of LiH-D mixed system [3].

\qquad As mentioned above the LiH$_{x}$D$_{1-x}$ mixed crystals at small x
value must have a two-mode character reconstruction of the optical phonons.
Moreover, according to the results of ref. 20 the observable situation with
the optical phonon rearrangement in LiH$_{x}$D$_{1-x}$ crystals satisfies
also the analytical criterion of the two-mode reconstruction at small
isotope concentration: 1/2W \TEXTsymbol{<} $\delta $ \TEXTsymbol{<} W. Here $%
\delta $ and W are the frequency shift of TO ($\Gamma $) phonon caused by
the isotopical defect and the band width of optical vibrations respectively
[27]. In Ref. 20 it is also noted that it is necessary to fulfil the more
rigid condition $\delta $ \TEXTsymbol{>} W to maintain the two-mode
behaviour of the phonon spectra up to an isotope concentration x $\simeq $
0.5.

The last result is obtained in a very crude approximation. Indeed, in Ref.
20 a model phonon spectrum was used for the calculation consisting of two
Gaussian bands and limited by $\omega _{TO}$ \TEXTsymbol{<} $\omega $ 
\TEXTsymbol{<} $\omega _{LO}$. Nevertheless it is necessary to note, that
this approximation does not greatly influence on the relation between W, and 
$\delta $,$\varepsilon _{0}$,$\varepsilon _{\infty }$, because the values
used in Ref. 20 were real (see, also Refs. 22,27). In Fig. 3 we plotted the
results from Fig. 1,2 as well as the data obtained for other crystals. From
Fig. 3b it is seen that two-mode optical phonons behavior is observed up to
x \TEXTsymbol{<} 0.4. The comparison of the optical phonon band width (W)
with the maximal shift ($\delta $) of the transverse optical phonon
frequency at the change of H to D shows [22] that the condition W 
\TEXTsymbol{>} $\delta $ is always fulfilled (see, also Ref. 27). That
relation contradicts the phonon CPA model prediction about two-mode optical
phonons rearrangement at x $\simeq $0.5. In our opinion this discrepancy is
due primarily to the strong isotope-induced additional phonon scattering
because the potential changes considerably on isotopic substitution in LiH
crystals. The observation of the local mode vibration at small x (see Fig.
1) may serve as further support for this conclusion. This conclusion agree
with the results of ref. 17. where it was emphasized that the observation of
the local mode vibration directly evidences the strong potential changes at
the isotope substitution. Therefore the weak phonon scattering approximation
in the CPA model is insufficient for the description of our experimental
results. On the other hand, the obtained conclusion quite agrees with the
fact that the parameter $\Delta \mu /\mu $ is not small for LiH-D system. As
it is well-known for Ge and C, the parameter $\Delta $M/M at the
isotopically substitution is extremely small (e.g. \TEXTsymbol{<}\TEXTsymbol{%
<}1) and the weak phonon scattering approximation of the CPA model is
sufficient to describe the experimental results in Refs. 17-19.

\qquad Now we are going to discuss briefly the dependence of the broadening
LO phonon lines on the isotope concetration (Fig.3a). The measured Raman
linewidth (see, also Fig. 4) is larger near the center of the composition
range than near the end points. it should be added that the width of the LO
phonon line depends not only on isotope concentration but also on the
excitation energy.

\bigskip

\bigskip Fig.4. Line shape of the excited light (1), 2LO line scattering at
4.2 K in LiH (2) and LiH$_{x}$D$_{1-x}$ mixed crystals ($\lambda _{excit}$ =
253.7 nm).

\bigskip

The results shown in Fig. 4 are the same as those obtained in earlier paper
[27] using uncoherent ultraviolet ($\lambda _{excit}$= 253.7 nm) radiation
for the excitation of the crystals. The observable addition structure on the
shortwavelength side of the 2LO ($\Gamma $) phonon replicas line in the
Raman spectra is due to the resonant excitation of the 2LO (L) phonons (for
details see Ref. 27 and references therein). Although the results depicted
in Fig. 3a are very similar to the results on diamond [18] it is necessary
to remark on the strong difference between our results on the two-mode
behavior of the LO($\Gamma $) phonons (at x \TEXTsymbol{<} 0.4) and one-mode
LO($\Gamma $) phonons in diamond. This difference, as well as the
observation of the local mode vibration (at small x) in the LiH$_{x}$D$%
_{1-x} $crystals, unambiguously allows us to make conclude about the strong
potential changes on the isotope substitution. Probably the nonlinear
dependence of the frequency shift and broadening LO phonon lines in the
Raman spectra on x may reflect the nonlinear character of the unharmonicity
[29].

\qquad I am grateful to Prof. F.F.Gavrilov for his generous supply with
samples used in these investigations, and Prof. E.E. Haller and Dr.
W.F.Banholzer for reprints of their papers, and Mr. D.Glover for
proof-reading the draft of this paper.

Figure captions

REFERENCES.

1. I.M. Lifshitz, Selected Papers, Science, Moscow, 1987 (in Russian).

2. A.A. Maradudin, E.W. Montrol, G.H. Weis and I.P. Ipatova, Theory of
Lattice Dynamics in Harmonic Approximation, Solid State Phys. Vol.3, edited
by F.Seitz, D.Turnbull and H.Erenreich (Academic, New York, 1971).

3. D.J. Montgomery and J.R. Hardy, J. Phys. Chem. Solids, Suppl. 1, 491
(1965).

4. M.H. Brodsky and E. Burstein, J.Phys. Chem. Solids, 28, 1655 (1967).

5. D. Laplaze, J. de Physique (Paris) 3, 1051 (1976).

6. J.L. Verble, J.L.Warren and J.L.Yarnell, Phys. Rev. 168, 980 (1968).

7. G. Wolfram, S.S.Jaswal and T.P.Sharma, Phys. Rev. Lett. 29, 160 (1972).

8. S.S. Jaswal, T.P.Sharma amd G.Wolfram, Solid State Commun. 11, 1151
(1972).

9. S.S. Jaswal, G.Wolfram and T.P.Sharma, J. Phys. Chem. Solids, 35, 571
(1974).

10. D. Laplaze, J. Phys. C: Solid State Phys. 10, 3499 (1977).

11. A. Anderson, F.Luty, Phys. Rev. B28, 3415 (1983).

12. V.I. Tyutyunnik, O.I.Tyutyunnik, Phys. Stat. Sol. (b)162, 597 (1990).

13. V.G. Plekhanov,V.A.Veltri, Sov. Phys. Solid State 33, 2384 (1991).

14. A.F. Kapustinsky, L.M.Shamovsky and K.S.Bayushkina Acta physicochim.
USSR 7, 799 (1937).

15. V.G. Plekhanov, in Proc. 20 Int. Conf. Phys. Semicond. edited by
E.M.Anastassakis and J.D. Joanopoules, Worid Science, Singapore, 1990, p.
1955.

16. V.F. Agekyan, V.M.Asnin, A.M.Kryukov, I.I.Markov, Sov. Phys. Solid State
31, 2082 (1989).

17. H.D. Fuchs, C.H.Grein, C.Tomsen, M.Cardona, W.L.Hansen, E.E.Haller and
K.Itoh, Phys. Rev. B43, 4835 (1991); P.Etchegoin, H.D.Fuchs, J. Weber,
M.Cardona, K.Itoh and E.E.Haller, ibid B48, 12661 (1993).

18. K.C. Hass, M.A.Tamor, T.R.Anthony and W.F. Banholzer, Phys. Rev. B45,
7171 (1992-1).

19. A.K. Ramdas, S.Rodriguez, M.Grimsditch, T.R. Anthony and W.F.Banholzer,
Phys. Rev. Lett. 71, 189 (1993); J.Spitzer, P.Etchegoin,M.Cardona,
T.R.Anthony and W.F.Banholzer,Solid State Commun. 88, 509 (1993).

20. R.J. Elliott, P.L.Leath, in Dynamical Properties of Solids, Vol. 2,
edited by G.K.Horton and A.A. Maradudin, North-Holland, Amsterdam,
1975,p.387.

21. D.W. Taylor, in Optical Properties of Mixed Crystals, edited by
R.J.Elliott and I.P.Ipatova North-Holland, Amsterdam-Oxford,1988, p.35.

22. V.G. Plekhanov, Opt. and Spectr. (St. Petersburgh) 76, 65 (1994).

23. N.R. Smyrl, E.L.Fuller and G.L.Powel, Appl. Spectr. 37, 38 (1983).

24. A.N. Babushkin, G.I.Pilipenko and F.F.Gavrilov, J.Phys. Condens. Matter
5, 8659 (1993).

25. V.G. Plekhanov, Solid State Commun. 76, 51 (1990).

26. Scott Kirkpatrick, Rev. Mod. Phys.45, 574 (1973).

27. Here W is the zone width, that is a difference between LO and TO phonons
and has the following expression: W = $\omega _{TO}\left( \varepsilon _{0}%
\text{ - }\varepsilon _{\infty }\right) /$ $\left( \varepsilon _{0}\text{ + }%
\varepsilon _{\infty }\right) $,where $\varepsilon _{0}$ and $\varepsilon
_{\infty }$ have the conventional meaning are equal to $\varepsilon _{0}$ =
12.9 and 14.0 and $\varepsilon _{\infty }$ = 3.61 and 3.63 for LiH and LiD
respectively .

28. V.G. Plekhanov,Phys. Lett. 148A,281 (1990)

29. A.A. Maradudin and S.Califano, Phys. Rev. B48, 12628 (1993).

\end{document}